\documentclass[conference,10pt,table]{IEEEtran}

\usepackage{lineno}
 
\usepackage{booktabs} 
\usepackage{balance} 
\usepackage{tabularx}
\usepackage{lipsum}

\usepackage{amsmath,amssymb,amsfonts}
\providecommand{\keywords}[1]{\textbf{\textit{Index terms-}} #1}
\usepackage{algorithmic}
\usepackage{graphicx}
\usepackage{subcaption}
\usepackage{textcomp}
\usepackage{fleqn}
\usepackage{soul}
\usepackage{color}
\usepackage{subcaption}
\usepackage{caption}
\usepackage{floatrow}%
\usepackage{wrapfig}
\usepackage{comment}
\usepackage{pifont}
\usepackage{tikz}
\usepackage{float}
\usepackage{amssymb}
\usepackage{floatrow}
\usepackage{cite}
\usepackage{xcolor}
\usepackage{lipsum}
\usepackage{enumitem}

\modulolinenumbers[5]
\usepackage{pdflscape}

\newcommand{\change}[1]{{\color{black}{#1}}}

\begin{document}
 
 


\change{
\title{OTFS-superimposed PRACH-aided Localization for UAV Safety Applications}
}

\author{\IEEEauthorblockA{Francesco Linsalata\IEEEauthorrefmark{1}, Antonio Albanese\IEEEauthorrefmark{2}\IEEEauthorrefmark{3}, Vincenzo Sciancalepore\IEEEauthorrefmark{2}, \\Francesca Roveda\IEEEauthorrefmark{1}, Maurizio Magarini\IEEEauthorrefmark{1}, Xavier Costa-Perez\IEEEauthorrefmark{2}\IEEEauthorrefmark{4}}

\IEEEauthorblockA{\IEEEauthorrefmark{1}Dipartimento di Elettronica, Informazione e Bioingegneria, Politecnico di Milano, Italy \\ 
\IEEEauthorrefmark{2}NEC Laboratories Europe GmbH, Heidelberg, Germany\\
\IEEEauthorrefmark{3} Departamento de Ingeniería Telemática, University Carlos III of Madrid, Legan\'es, Spain\\
\IEEEauthorrefmark{4} i2cat Foundation and ICREA, Barcelona, Spain\\
}}

\maketitle

\begin{abstract}
The adoption of Unmanned Aerial Vehicles (UAVs) for public safety applications has skyrocketed in the last years. Leveraging on Physical Random Access Channel (PRACH) preambles, in this paper we pioneer a novel localization technique for UAVs equipped with cellular base stations used in emergency scenarios. We exploit the new concept of Orthogonal Time Frequency Space (OTFS) modulation (tolerant to channel Doppler spread caused by UAVs motion) to build a fully standards-compliant OTFS-modulated PRACH transmission and reception scheme able to perform time-of-arrival (ToA) measurements. First, we analyze such novel ToA ranging technique, both analytically and numerically, to accurately and iteratively derive the distance between localized users and the points traversed by the UAV along its trajectory. Then, we determine the optimal UAV speed as a trade-off between the accuracy of the ranging technique and the power needed by the UAV to reach and keep its speed during emergency operations. Finally, we demonstrate that our solution outperforms standard PRACH-based localization techniques in terms of Root Mean Square Error (RMSE) by about $20\%$ in quasi-static conditions and up to $80\%$ in high-mobility conditions. 
\end{abstract}


\section{Introduction}

Cellular-aided localization is becoming increasingly important as demonstrated by the design of next generation mobile networks~\cite{6g_whitepaper}. With the growing accuracy of mobile network-based positioning, telco operators may bootstrap novel business opportunities by taking advantage of the real-time knowledge of their user positions and deliver value-added services such as goods tracking (in manufacturing and logistic operations) or vehicle navigation (within the ever-growing trend of autonomous driving). This will heavily impact the  5G and 6G markets enabling new sources of revenues for service providers~\cite{Giordani2019}.

Additionally, location information may be further disclosed to public authorities so as to support first responders in search-and-rescue operations in the aftermath of natural or human-made disasters~\cite{bartoletti2018}. In critical operation conditions, public authorities may benefit from the higher reliability provided by network-based localization solutions with respect to Global Navigation Satellite Systems (GNSSs) for the following reasons. First, while the GNSS technology might achieve a localization accuracy of about $3 \, \text{m}$, it suffers from shadowing due to tree foliage, debris or multipath in urban canyons, thus impairing its effectiveness in Non Line-of-Sight (NLoS) conditions.
Second, on top of such limitations, there is no effective way to obtain remote access to users GNSS information through the network, due to the lack of ad-hoc Application Programming Interfaces (APIs) in the 3rd Generation Partnership Project (3GPP) standard guidelines for that purpose~\cite{ETSIreq}.

Moreover, terrestrial cellular infrastructure may be partially or totally disrupted in emergency situations, thus calling for effective alternative solutions that 
make victim localization a reliable process~\cite{Namuduri2017}. Unmanned Aerial Vehicles (UAVs) keep drawing the interest of the scientific community for their capability of operating in difficult-to-reach locations thereby giving birth to fast and agile network deployments. Specifically, UAV-equipped base stations may be effectively employed in search-and-rescue operations---without requiring network infrastructure support---due to their ability for  detecting victims through their cellular devices~\cite{albanese2021}. However, UAV scenarios represent a high-mobility case suffering from high Doppler spread that, in turn, may hinder the air-to-ground channel quality and reduce the obtained ranging measurement accuracy.  

The Doppler spread issues with Orthogonal Frequency-Division Multiplexing (OFDM) modulation have been exhaustively evaluated in the literature. In~\cite{wenbo}, via a railway testing campaign replicating a high-mobility propagation scenario,  the authors demonstrated the impact of speed and Doppler spread on the user signal quality, i.e. how UEs may experience network accessing problems due to compromised orthogonality on the Physical Random Access Channel (PRACH).
In~\cite{micheli}, the Doppler shift problem is also considered by proposing an improved two-step PRACH preamble detection method for high-speed scenarios. This allows mitigating the degradation introduced by the channel at a high-complexity cost with respect to the standard single-step detection.


In this context, a novel modulation scheme may come to the rescue: Orthogonal Time Frequency Space (OTFS) has substantially attracted industrial and academic interest as an effective solution to efficiently cope with the time-variant multipath channel experienced by OFDM communication when the terminals speed is up to $500$ km/h, as expected in the future 6G use-case scenarios. OTFS spreads the information symbol waveform over the entire delay-Doppler plane, achieving near-constant channel gain even in channels with high Doppler spread~\cite{Hadani2017b,Liu2021}. \change{It is well known that OTFS outperforms OFDM over doubly-selective channels. Even though processing consecutive bursts of OFDM symbols offers a possible workaround, this would increase the latency in contrast to the low-latency requirements envisioned for 6G.}

In this paper, \change{leveraging the OTFS sensing capabilities in the delay-Doppler domain}, we propose a novel and fully 3GPP-compliant OTFS-based emergency localization technique that can be on-boarded on a UAV able to scan the disaster area and accurately retrieve victim positions, even in high-mobility scenarios (e.g., people fleeing from an imminent danger). Our novel solution appears \emph{the first of its kind} as it is suitable for upcoming 6G networks design supporting emergency situations by making use of the initial PRACH procedure as a guaranteed way to spot and reach people in disaster areas.
The contributions of this paper can be summarized as follows: $i$) we introduce an OTFS-modulated PRACH procedure that is fully-compliant with the current 5G-NR standard, $ii$) we design a ranging technique leveraging on the delay estimate of an OTFS-modulated PRACH; $iii$) we derive high-mobility air-to-ground channels via ray-tracing simulations and $iv$) we exhaustively evaluate our proposed ranging localization solution with realistic-channel assumptions in case of snowfall.

The remainder of this paper is the following. Section~\ref{sec:preliminaries} provides an overview of the OTFS modulation and the 5G-NR PRACH. Section~\ref{sec:model} introduces our solution leveraging on OTFS modulation to perform Doppler-tolerant localization over the 5G-NR PRACH. Section~\ref{sec:scenario} describes the simulation settings and scenario, whereas Section~\ref{sec:perf_eval} validates the overall technique through both analytical derivations and MATLAB simulations performed on realistic air-to-ground channels, and investigates the optimal UAV speed for the task. Lastly, Section~\ref{sec:conclusions} concludes the paper.

\section{Preliminaries}
\label{sec:preliminaries}

We first introduce the general OTFS scheme preliminaries. We then describe the theoretical characteristics and practical implementations of the 5G-NR PRACH procedure while shedding the light on the positive impact of 5G numerology on the theoretical localization accuracy.

\subsection{OTFS Modulation}
The main idea behind the OTFS modulation is to spread the basis waveform over the whole time-frequency plane, thereby achieving high diversity order and resilience against Doppler spread. Specifically, OTFS operates in the delay-Doppler domain, transforming an Orthogonal Frequency-Division Multiplexing (OFDM) fading time-variant channel into a non-fading time-invariant channel~\cite{Hadani2017a}. To this aim, the information symbols $x[l,k]$ produced in the delay-Doppler domain are mapped onto the time-frequency domain as $X[n,m]=\text{ISFFT}\{x[l,k]\}$, with ISFFT indicating the Inverse Symplectic Finite Fourier Transform, yelding 
\begin{align} \label{Xnm}
    X[n,m] =\frac{1}{MN}\sum\limits_{l,k} x[l,k] e^{j2\pi(\frac{nk}{N}-\frac{ml}{M})},
\end{align}
where $M$ represents the number of subcarriers and $N$ is the number of symbols per subcarrier. Through the ISFFT, the delay-Doppler symbols are converted into a linear combination of basis functions, namely two-dimensional orthogonal complex exponential functions, covering the entire time-frequency domain. 
\change{The continuous time signal is obtained by applying the so-called Heisenberg Transform to $X[n,m]$ in (\ref{Xnm}) as}
\begin{align}
   s(t)=\sum\limits_{m=0}^{M-1} \sum\limits_{n=0}^{N-1} X[n,m]g_{tx}(t-nT) e^{j2\pi m\Delta f(t-nT)}, 
\end{align} 
where $g_{tx}(t)$ is the pulse shaping filter, $\Delta f$ is the subcarriers spacing, and $T$ is the symbol interval. The received signal $r(t)$ is the convolution between the transmitted signal $s(t)$ and the time-variant channel impulse response,
\change{namely
\begin{align} \label{eq:CT_received_signal_singleant}
 r(t) & = \int h(\tau,t)\, x(t-\tau) \,d\tau  + n(t) = \nonumber\\
      &=\iint h(\tau,\nu) \, e^{j 2 \pi \nu (t - \tau)} x(t-\tau) \, \mathrm{d} \nu \, \mathrm{d} \tau + n(t),
\end{align}
where $h(\tau,\upsilon)$ is dubbed as Delay–Doppler spreading function and $h(\tau,t)$ is the time-variant channel response, which can be shown to be equivalent to
\begin{align}\label{eq:channel_model}
    h(\tau,t) = \sum_{i=1}^I \alpha_i \, e^{j 2 \pi \nu_i t} \, \delta(\tau-\tau_i),
\end{align}
where $\alpha_i$ is the amplitude of the $i$-th path, depending on the UAV-target distance (range), $\tau_i$ and $\nu_i$ are the delay and the Doppler shift of the $i$-th path, respectively.
}
At the receive side, the Wigner transformation is first applied~\cite{Hadani2017b} as
\begin{equation} \label{ytf}
 Y(t,f)=\int g_{rx}(t' -t) \cdot r(t')e^{-j2\pi f(t'-t)}\,dt',
\end{equation}
such that $Y[n,m]=Y(t,f)|_{t=N,f=m\Delta f}$.
Then, the delay-Doppler domain signal is retrieved via Symplectic Finite Fourier Transform (SFFT) \change{as
\begin{align} \label{eq:DDreceivedsignal}
    Y[l,k]=  \frac{1}{\sqrt{N M}} \sum_{n=0}^{N-1} \sum_{m=0}^{M-1} Y[m,n] e^{j 2 \pi ( \frac{nk}{N}-\frac{ml}{M})}.
\end{align}}

\subsection{PRACH in 5G-NR}
\label{s:prach}
According to 3GPP guidelines (see, e.g., ~\cite{ETSIreq}), a User Equipment (UE) undergoes the Random Access Procedure (RAP) whenever it initiates a connection to a base station (BS), e.g., initial  access  to  the  network  or  handovers\footnote{\change{The BS regularly transmits the essential parameters for initial access in the Master Information Block (MIB) over the Physical Broadcast Channel (PBCH).}}. Such logical procedure is performed over the Physical Random Access Channel (PRACH), which ensures the simultaneous connection between multiple users and one base station (BS). Through the transmission of a specific preamble sequence, a UE willing to synchronize with the BS notifies its presence.


\textbf{Zadoff-Chu Sequences.}
Zadoff-Chu (ZC) sequences are non-binary sequences of constant amplitude that satisfy the Constant-Amplitude Zero Auto-Correlation (CAZAC) property~\cite{Sesia2009}.
This kind of signal is transmitted over the PRACH for timing, frequency and symbol synchronization.
The general form of a ZC sequence is
\begin{align}
    a_{q}(n)=\exp\left(-j2\pi q \frac{n(n+1)/2}{N_{ZC}}\right),
\end{align}
where $N_{ZC}$ is the length of the sequence, $q$\,$=$\,$1,...,N_{ZC}$ is the root index, and $n$\,$=$\,$0,...,N_{ZC}-1$ is any integer number. 
CAZAC sequences hold two fundamental properties: \textit{i)} a ZC sequence is invariant to Fast Fourier Transform (FFT) or Inverse Fast Fourier Transform (IFFT) operations; \textit{ii)} circularly shifted copies of the sequence are uncorrelated for a fixed root~\cite{LINSALATA}.

\textbf{PRACH Preamble Detection.} 
The transmission of the PRACH preamble is implemented in the frequency domain~\cite{ETSIreq}, applying the Discrete Fourier Transform (DFT) to a ZC sequence of length $N_{ZC}$ (equal to $139$ or $839$ depending on the use of short or long preamble, respecitvely). 
At the receive side, the process is reversed in order to obtain 
a sequence that can be compared to the original ZC root sequence~\cite{LINSALATA}.
The PRACH receiver can benefit from the ZC format and the CAZAC property by computing the received Power Delay Profile (PDP) as
\change{\begin{align}
  \text{PDP}(l)=|z_{u}(l)|^2=\Bigg|\sum_{n=0}^{N_{ZC}-1} y(n)x^*_{u}[(n+l)] \Bigg|^2, 
\end{align}}
where $z_{u}(l)$ is the discrete correlation function at lag $l$, $y(n)$ is the post processed received sequence, $x_{u}(n)$ is the ZC root sequence repeated at the receiver. and $(\cdot)^*$ denotes the complex conjugate operation.

\textbf{Time-of-Arrival Evaluation.} 
The detection of the preamble is used by the BS to evaluate the Time-of-Arrival (ToA) of the channel~\cite{LINSALATA}.
This operation is implemented by computing a power detection threshold $T_{total}$---the power level below which the received signal is supposed to be only noise---as $T_{total}=T_{TH} \cdot m_{tot}$, where $T_{TH}$ is the optimal threshold evaluated from the theory~\cite{Sesia2009} to satisfy a given false alarm probability, and  $m_{tot}$ is the average power of correlation PDP. The time sample $k$ at which the PDP is maximum (and greater than $T_{total}$) is the sample delay associated with the user. 
In our solution, we evaluate the physical three-dimensional distance between the user and the UAV as $d_{ext}= k \cdot \Delta t \cdot c$,
where $\Delta t$ is the system sampling time and $c$ the speed of light. The time resolution $\Delta t$ is proportional to the length of the DFT $N_{FFT}$ and $\Delta f$ and can be written as $\Delta t=\frac{1}{\Delta f \cdot N_{DFT}}$ so that, the lower the time resolution, the higher the precision in the evaluation of the distance.
It is worth noting that the choice of $N_{DFT}$ and $\Delta f$ is related to the physical implementation of both the transmitting and receiving system, as explained below. By enabling subcarrier spacing $\Delta f$ tuning, 5G numerology can be exploited to achieve higher localization accuracy while keeping full compatibility with the proposed OTFS design.


\section{Proposed System Design}
\label{sec:model}

\begin{figure*}[t!]
     \vspace{2mm}
     \centering
     \begin{subfigure}{0.47\textwidth}
         \centering
         \includegraphics[width=\textwidth]{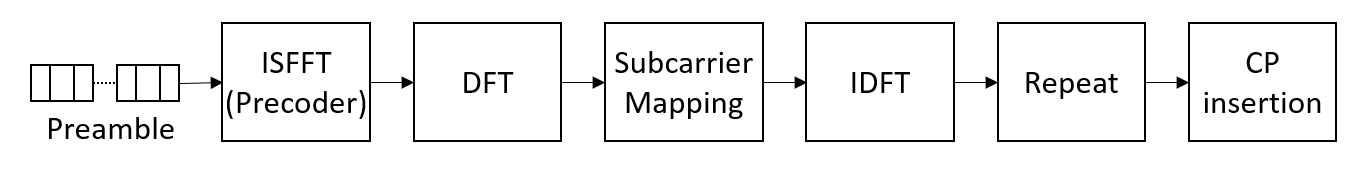}
         \caption{Transmitter}
         \label{fig:sub1}
     \end{subfigure} \hspace{.73cm}
     \begin{subfigure}{0.47\textwidth}
         \centering
\includegraphics[width=\textwidth]{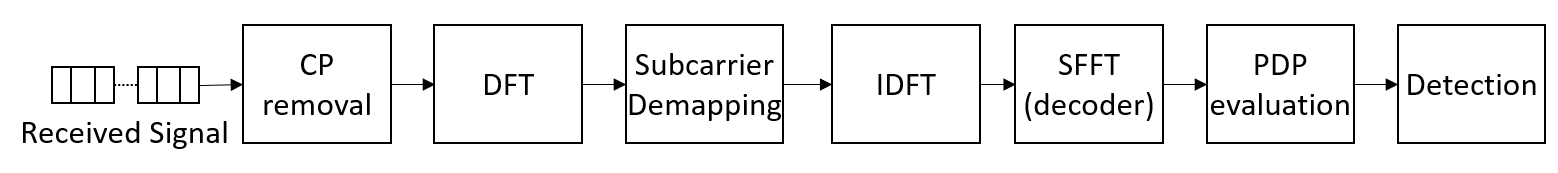}
         \caption{\change{Receiver}}
         \label{fig:sub2}
     \end{subfigure}
        \caption{Proposed OTFS-based PRACH Architecture}
    \label{fig:PRACH_OTFS}
\end{figure*}

The detection of the PRACH preamble and the ToA estimation can be used not only for synchronization purposes, but also for target UE localization in emergency situations when GNSS information is not available. \change{In these cases, the mobile network can keep rejecting UE attachments with a proper sequence of Mobility Management Entity (MME) response messages such that a RAP time-out is never reached~\cite{albanese2021}.} 

Nevertheless, as the current 5G-NR standard is not effective against Doppler-impaired and time-variant channels, we introduce an OTFS-based ZC precoding and decoding scheme able to both mitigate and take advantage of such hard channel conditions.
Our proposed transmission and reception schemes are respectively sketched in Fig.~\ref{fig:PRACH_OTFS}. The main body of the block diagram is 5G-NR standards-compliant, allowing adapting the OTFS-based model to the already existing architectures based on the OFDM modulation. We introduce a novel precoder that implements the Inverse-SFFT, and a decoder at the receiver side, which performs the SFFT to obtain the initial preamble. The OTFS modulation allows compensating for the Doppler spread due to to the mobility of the UAV-mounted BS, thus improving the ToA estimation.
In the proposed scheme, we first generate the preamble using a ZC sequence \change{that is repeated and mapped in each row of a 2D matrix $X \in \mathbb{C}^{N_{ZC} \times M}$}. Before the conventional OFDM modulator operations, we apply the ISFFT to achieve high diversity against Doppler spread as per Eq.~\eqref{Xnm}. After the Fourier transform, we map the preamble onto the subcarrier basis and transmit it in the time domain after adding a Cyclic Prefix to mitigate the inter-carrier interference and guarantee the orthogonality of the PRACH sequence regardless of delay and Doppler spread. 

%
%
After removing the CP, the receiver performs subcarrier demapping and the SFFT as per Eq.~\eqref{ytf} to obtain the received preamble. Then, the estimated ToA is derived through the cross-convolution between the received preamble and a copy of the original transmitted Zadoff-Chu sequence. The peak position of the convolution, identified as $k$, is the sample delay between the two signals, which yields $d_{ext}=\frac{c}{\Delta f \cdot N_{DFT}}k$, as by Section~\ref{s:prach}.
\change{Then, the collected ToA measurements can be fed to well-known localization techniques (e.g. multilateration) to target the user~\cite{albanese2021}.}

\begin{figure*}
     \centering
     \begin{subfigure}[t]{0.44\textwidth}
         \centering
         \includegraphics[width=\textwidth]{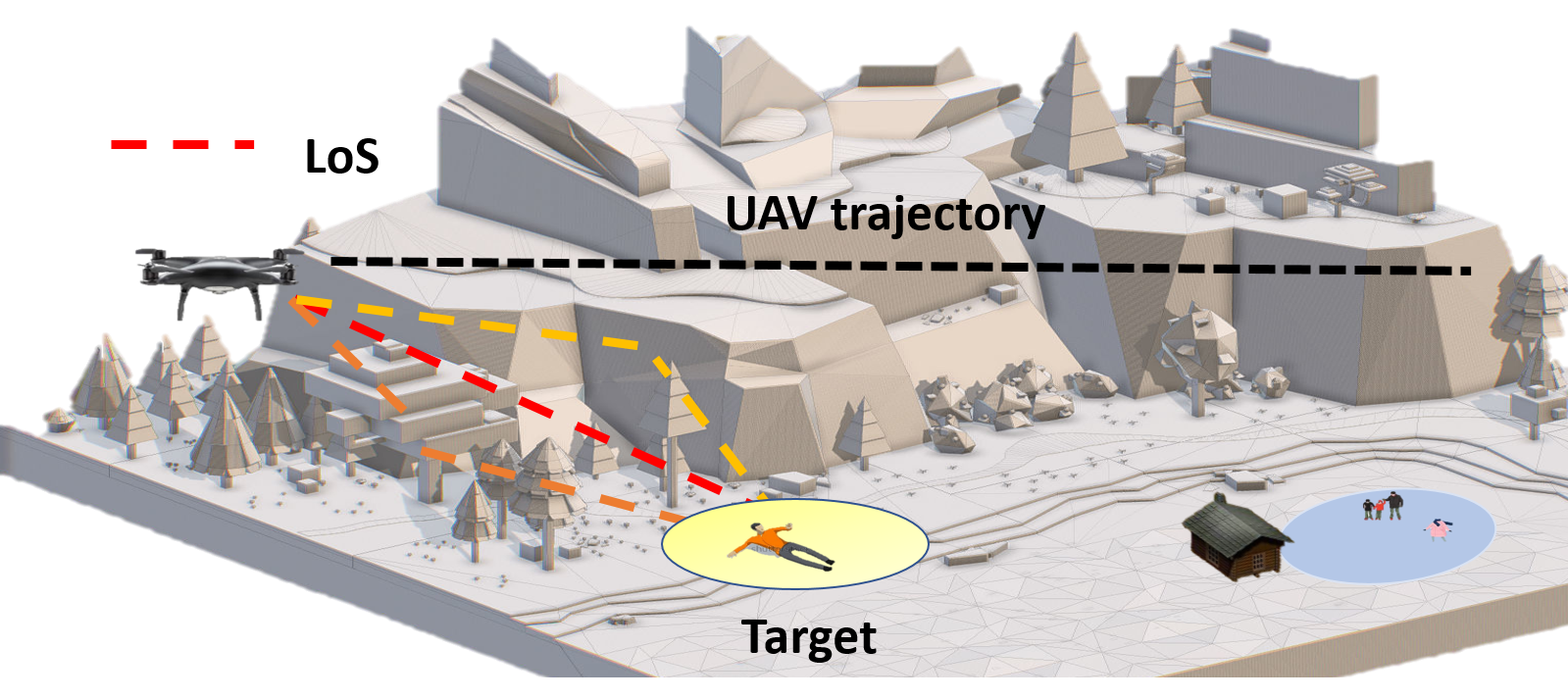}
         \caption{Realistic scenario}
         \label{fig:real scenario}
     \end{subfigure} \hspace{1.1cm}
     \begin{subfigure}[t]{0.44\textwidth}
         \centering
\includegraphics[width=\textwidth]{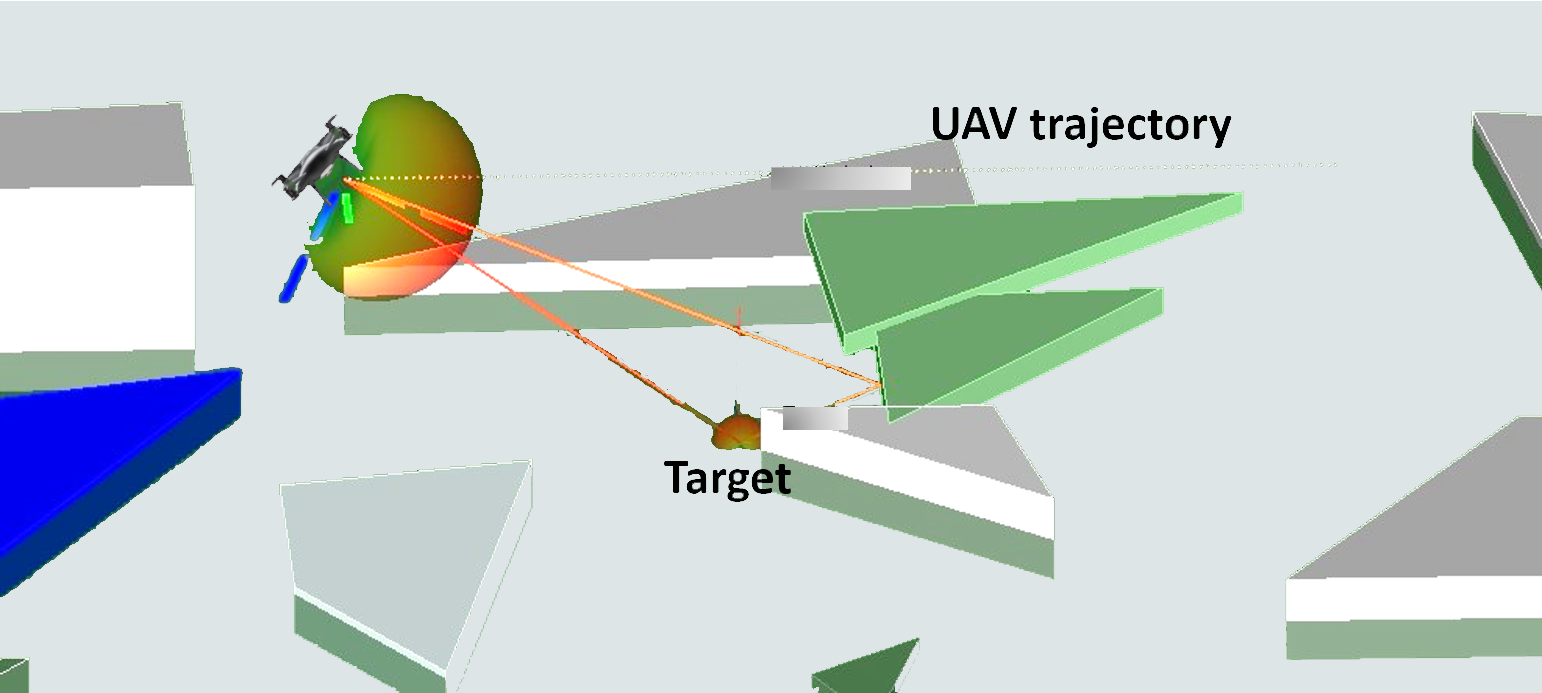}
         \caption{Wireless InSite model}
         \label{fig:drone trajectoryWI}
     \end{subfigure}
        \caption{UAV emergency localization operations using a commercial ray-tracing simulator.}
        \label{fig:drone rtajectory}
\end{figure*}

\section{Simulations Setting \& Scenario }
\label{sec:scenario}

The parameters used in the OTFS-based PRACH implementation are 5G-NR standard compliant, according to the specifications in~\cite{ETSIreq,Keysight2017} and listed in Table~\ref{tab:parameters for the evaluation}. 
\begin{table}[b!]
\caption{Simulation settings}
\label{tab:parameters for the evaluation}
\centering
\resizebox{.65\linewidth}{!}{%
\begin{tabular}{cc|cc}
\textbf{Parameter} & \textbf{Value} & \textbf{Parameter} & \textbf{Value}\\  
\hline
\rowcolor[HTML]{EFEFEF}
$N_{ZC}$ & 139     & $N_{DFT}$  &  2048 \\
$q$           & 1      & $M$              & 1024  \\
\rowcolor[HTML]{EFEFEF}
 $P_t$ & 23 dBm & $G_{tmax}$ & 10 dB \\
\end{tabular}%
}
\end{table}

\subsection{Channel Model}


Aiming at realistic simulations, we make use of Wireless InSite~\cite{guidaWI} to \change{extract} the channel \change{parameters} \change{as per Eq.~\eqref{eq:channel_model}}. 
The realistic scenario is sketched in Fig.~\ref{fig:real scenario}, while Fig.~\ref{fig:drone trajectoryWI} depicts the corresponding Wireless InSite simulation scenario. Each block corresponds to an obstacle  (i.e. stones, trees, and small buildings debris) whose density and electromagnetic properties are defined accordingly. All the objects and the rest of the scenario are covered by a layer of snow, which has a variable depth between $0$ and $3\, \text{m}$, and some rubble located in the proximity of the UE and made of different materials, one per object face. The UAV-mounted BS receives a sinusoidal signal at $1775$ MHz (LTE Band 3, Earfcn UL 19850) from a UE that is stuck under a snow cover in a static and unknown position. The UAV follows a linear trajectory over the UE, with different velocities, sampled every $0.5\, \text{m}$, up to a total of 140 positions. 
We define $|h|^2$ as the channel gain, which can be evaluated as $|h|^2=10^{\frac{Pg}{10}}$ with $Pg = P_{r}-P_{t}-G_{rmax}-G_{tmax} + LS\text{ [dB]}$,
where $G_{rmax}$ and $G_{tmax}$ are the maximum antenna gains, LS is the additional loss, \(P_{r}\) is the received power 
\begin{align}
    Pr=P_{t}\bigg{(}\frac{\lambda}{4\pi d}\bigg{)}^2 [G_{t}(\theta)G_{r}(\theta)]^2,
\end{align}
\(G_{t}(\theta)\) is the gain of transmitting omnidirectional antenna and \(G_{r}(\theta)\) is the receiving directional antenna gain that depends on both the angle between the transmitter and the receiver, and the tilt angle as described below. 

\subsection{On-board Antenna}
The gain of the UAV-mounted BS antenna $G_{r}$ is directly related to the tilt angle of the UAV as~\cite{Dandanov2017}
\begin{align}
\begin{split}
G_{r}(\gamma,\theta,\theta_{v})&=A_{T}(\gamma,\theta,\theta_{v}) +G_{r.max},\\
A_{T}(\gamma,\theta,\theta_{v})&=A_{H}(\gamma,\theta,\theta_{v})+A_{V}(\gamma,\theta,\theta_{v}),
\end{split}
\label{eq:gain directional antenna}
\end{align}
where  \(A_{H}(\gamma,\theta,\theta_{v})\)\,$=\hspace{-0.41cm}-12(\frac{\gamma}{\gamma_{3dB}})^2$, and \(A_{V}(\gamma,\theta,\theta_{v})\)$=\hspace{-0.1cm}$\,$-12(\frac{\theta-\theta_{v}}{\theta_{3dB}})^2$ are horizontal and vertical antenna patterns, while \(\gamma_{\text{3dB}}\) and \(\theta_{\text{3dB}}\) represent horizontal and vertical 3 dB bandwidths, respectively. Variables \(\gamma\) and \(\theta\) are the angles between the normal to the antenna and the direction of measurements in the horizontal and vertical planes. The higher is the antenna tilt, the higher is the match, therefore the gain of the receiving directional antenna.
To start moving and then increasing its speed \(v_{x}\), the UAV needs to tilt itself with an angle \(\theta_{v}\), which lays between the UAV tilt and the horizontal plane.
In~\cite{Okada2019}, the relationship between the tilt angle and the speed is derived by calculating the pitch angle $\theta_{\xi}$, which stays between the x-axis and the yaw axis of the UAV, as
\begin{align}
\begin{small}
    \hspace{-0.2cm}\theta_{\xi}=cos^{-1}\Bigg\{\sqrt{\left(\frac{mg}{\rho C_{D}^{\epsilon}A_{\epsilon}v_{x}^2}\right)^2+1}-\left(\frac{mg}{\rho C_{D}^{\epsilon}A_{\epsilon}v_{x}}\right)\Bigg\},  \label{eq:moving angle1}
\end{small}
\end{align}
where $\theta_{v}$\,$=$\,$90^{\circ}$\,$-$\,$\theta_{\xi}$, \(m\) is the weight of the UAV, \(g\) is the gravitational acceleration, \(\rho\) is the air density, \(C_{D}^{\epsilon}\) is the resistance coefficient, and \(A_{\epsilon}\) is the swept area.
This UAV feature can be used to improve the ranging accuracy by selecting the most suited on-board directional antenna according to its radiation pattern. Specifically, in a static position the main lobe is parallel to the ground while, increasing the speed, the main lobe tilts down defining the angle $\theta_{v}$. The highest overall gain is reached when the antenna main lobe direction is pointed exactly towards the user, namely for $\theta = \theta_{v}$. 

\section{Performance Evaluation} \label{sec:perf_eval}

Hereafter, we illustrate the performance improvements in terms of distance estimation accuracy of the OTFS-based PRACH preamble scheme, as proposed in Fig.~\ref{fig:PRACH_OTFS}, with respect to the current 5G-NR standard, based on OFDM modulation. 

\begin{figure}[t!]
 \vspace{0.2cm}
 \centering
    \includegraphics[width=0.97\textwidth]{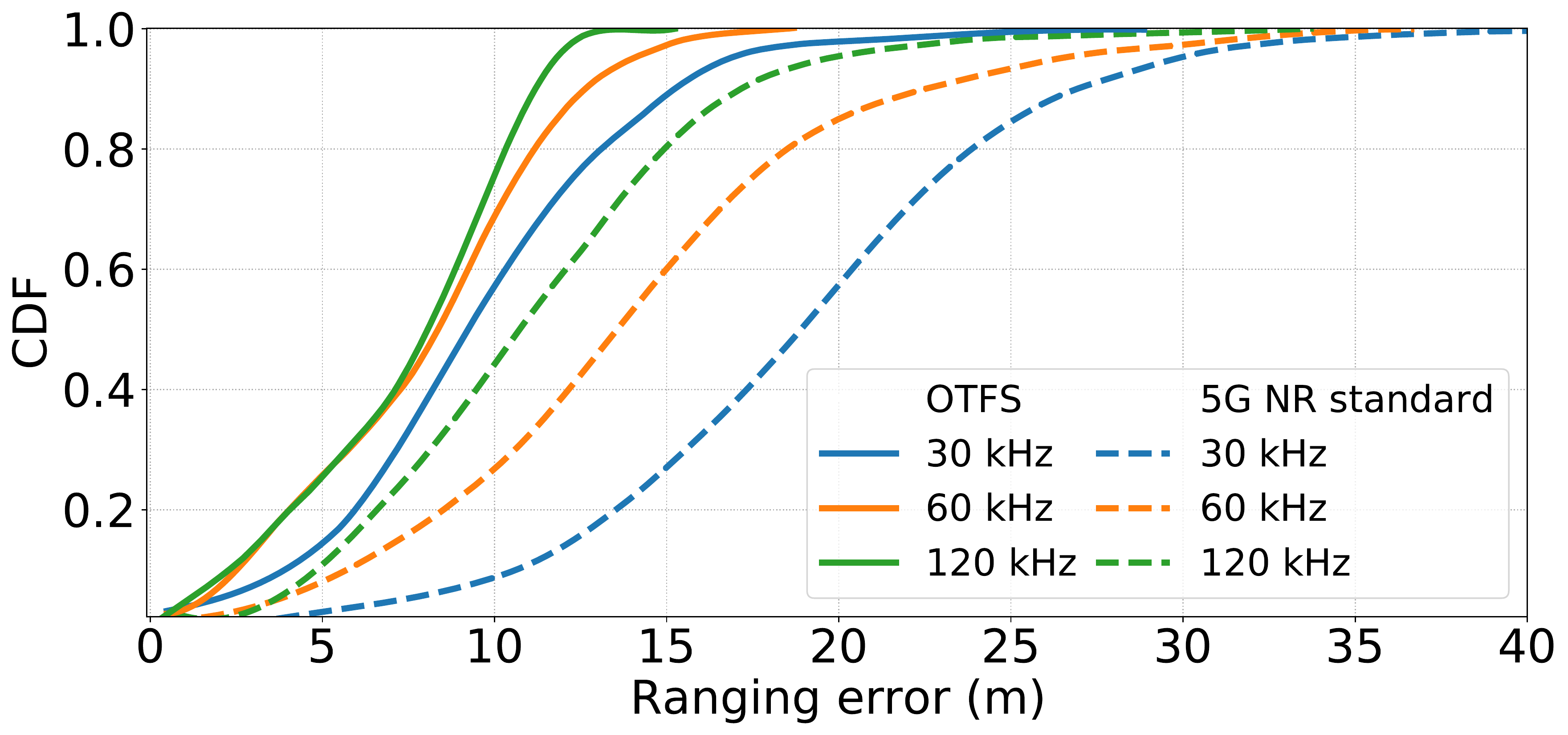}
    \caption{Cumulative Distribution Function of ranging error for different subcarrier spacing values.}
    \label{fig:CDF}
\end{figure}

Fig.~\ref{fig:CDF} shows the cumulative distribution function (CDF) of the absolute ranging error in NLoS conditions for a UAV moving at $10$~m/s equipped with an omnidirectional antenna against different subcarrier spacings obtained by, e.g., 5G numerology. In this scenario, OTFS modulation outmatches the current OFDM 5G NR standard by at least $5$~m (on average) with any subcarrier spacing configuration. Moreover, the error decreases as the subcarrier spacing increases since a \(\Delta f\) rise leads to a \(\Delta t\) reduction, which, in turn, yields higher ranging resolution. 

\textbf{Line-of-Sight conditions.} We evaluate the localization accuracy by means of the Root Mean Square Error (RMSE), defined as $\sqrt{\frac{1}{N}\sum_{n}\epsilon^2(n)}$, where $\epsilon(n) = d(n)-d_{ext}(n)$ is the 3D distance error for each position of the UAV trajectory. In LoS conditions, the RMSE can be directly related to the OTFS parameters as per~\cite{Gaudio2019},
\begin{align}
\label{eq:rmse_model}
\small
    \mathrm{RMSE}_{LoS}=\sqrt{\frac{6}{4\pi MK(M^{2}-1)P_{t}}\cdot \frac{1}{|h|^2}},
\end{align}
where $M$ is the number of subcarriers, $K$ the number of symbols, and \(P_{t}\) is the mean transmitted power. 
Thanks to the favourable LoS conditions allowing a high enough received Signal-to-Noise Ratio (SNR), the peak of the cross-correlation between the received PRACH preamble and a local copy stored at the receiver is likely detectable regardless of the UAV speed. 


\textbf{Non Line-of-Sight conditions.} \label{subsec:NLos} In presence of multipath fading, the error derives from both the LoS link and the (multiple) NLoS links. Moreover, in such conditions, UAV physical configurations such as the antenna tilt play a major role into the ranging measurement process. The estimated distance obtained through the Time of Arrival (TOA) measurements can be written as $d_{ext}=d+\epsilon_{n}$, with $\epsilon_{n}=\epsilon_{n,LoS}+\gamma\epsilon_{n,NLoS}$,
\change{where $\epsilon_{n,LoS}$ and $\epsilon_{n,NLoS}$ refer to the error components purely related to LoS and NLoS propagation, respectively. To reflect this in the overall RMSE definition, we set parameter $\gamma$ to $0$ in LoS conditions, while we set it to $1$ in NLoS cases. 

It is worth pointing out that we identify as LoS points the trajectory points at which the UAV experiences only a direct communication link to the UE as well as those at which the received power at the UAV side from the highest NLoS component is negligible with respect to the LoS one. Along these lines, NLoS points are defined as the trajectory points at which the received NLoS components have a comparable power with the LoS path, thus affecting the accuracy of the ToA measurements. 

Although \(\epsilon_{n,LoS}\) depends purely on the distance resolution \(\Delta d\) as the real measurements are not necessarily integer multiples of the sampling distance, \(\epsilon_{n,NLoS}\) is also affected by the distribution of path-length of first arriving NLoS signals (\(d_{fp}\)) and their difference with respect to an integer multiple of the sampling distance of the PRACH-based ranging solution.
}

Indeed, the RMSE of the ranging measurements is defined as
\begin{align} \label{rmsenlos}
   \sqrt{E[(\epsilon_{n})^2}]&= 
\sqrt{\frac{1}{N}\sum_{n}(\epsilon_{n,LoS}+\gamma\epsilon_{n,NLoS})^2}, 
\end{align}
where $N$ is the total number of UAV positions along the trajectory. 


\begin{figure}[t!]
    \centering
 {\includegraphics[width=\textwidth]{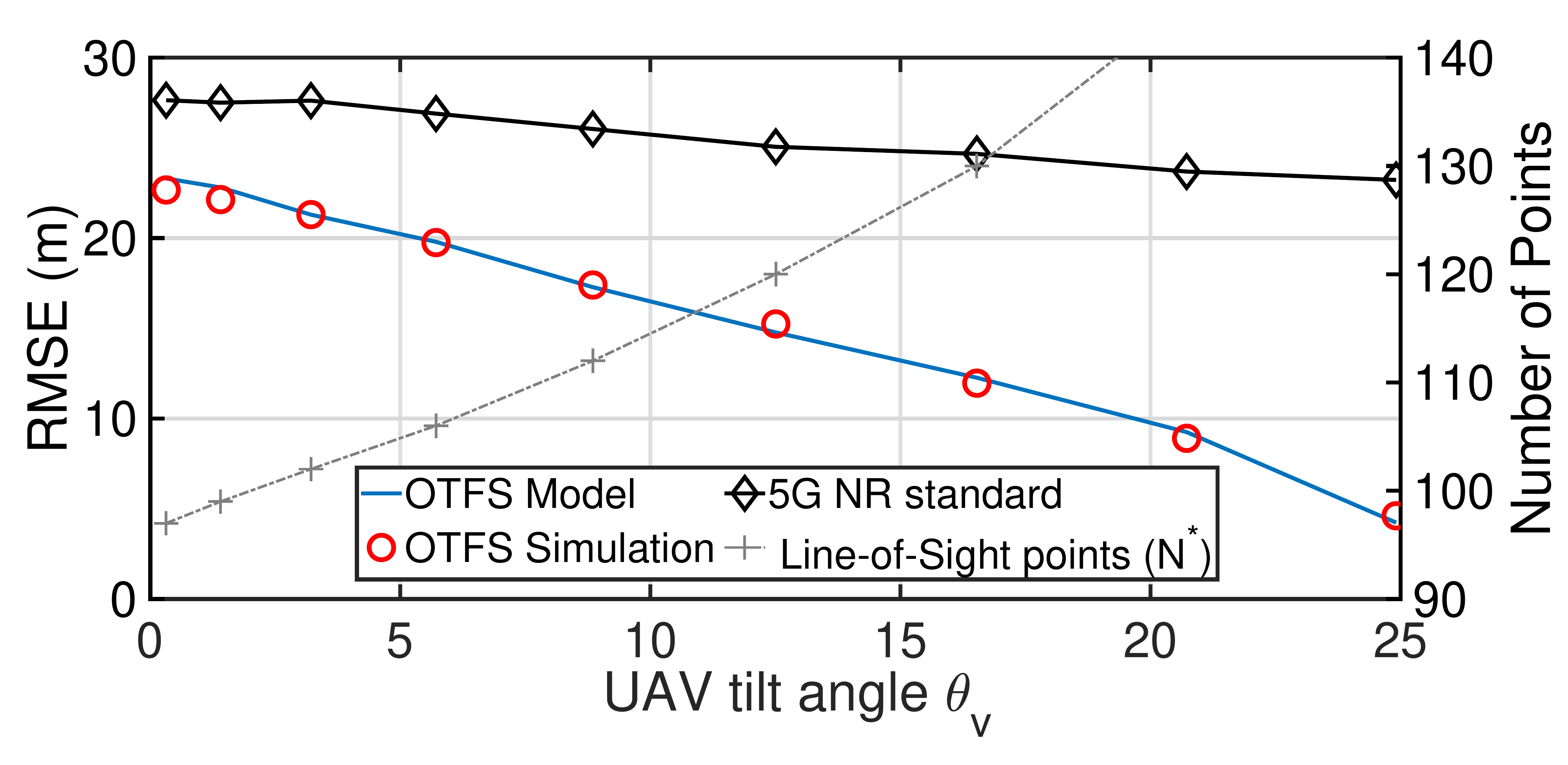}}
    \caption{RMSE performance and analysis of LoS occurrences (in terms of trajectory points) versus UAV antenna tilt.}
    \label{fig:RMSE vs theta}
\end{figure}

\textbf{UAV Antenna tilt effect.} We prove that the larger the tilt and the smaller the angle between the main path and the antenna direction \(\theta_{a}\), the higher the number $N_{LoS}$ of LoS points.  It is well-known that LoS condition holds until when \(\theta_{a}\) is less than or equal to the First Null Beamwidth (FNB) of the antenna, i.e.,
\begin{align}
\label{eq:FNB}
    \theta_{a}=180-\theta-\theta_{v},\,\, \theta_{a}\leq FNB.
\end{align}
The trajectory point index $n$ at a given distance $d$ from the target can be found as
\begin{align}
    n=p_{0}+\frac{\sqrt{d^{2}-h^{2}}}{dp},
\end{align}
where 
\(p_{0}\) is the index of the point at which the UAV is right over the victim, i.e. such that \(\theta =\arctan(\frac{h}{\sqrt{d^{2}-h^{2}}})\) is equal to zero, while \(dp\) is the distance between two successive points of the trajectory. The last LoS point can be obtained by setting $\theta = \arctan(\frac{h}{\sqrt{d_{max}^{2}-h^{2}}})$ in Eq.~\eqref{eq:FNB}, leading to
\begin{align}
    N_{LoS}=\frac{h}{dp\cdot \tan(180-FNB-\theta_{v})}+p_{0}.
\end{align}
\change{Note that we filter such points by removing those at which a strong multi-path instance corrupts the ToA estimation, thus obtaining \(N^{\ast}\) empirically validated LoS points.} In Fig.~\ref{fig:RMSE vs theta}, we depict the RMSE performances against the UAV antenna tilt angle. Although both OTFS and OFDM modulations provide better accuracy while the UAV moves (due to the higher number of LoS points traversed by the UAV along its trajectory), OTFS achieves higher performance gains by taking advantage of the UAV speed in the delay-Doppler domain. Indeed, the standard OFDM scheme shows a constant degradation (more than 50\% with respect to the OTFS solution). 

\begin{figure}[t!]
  \includegraphics[width=.97\textwidth]{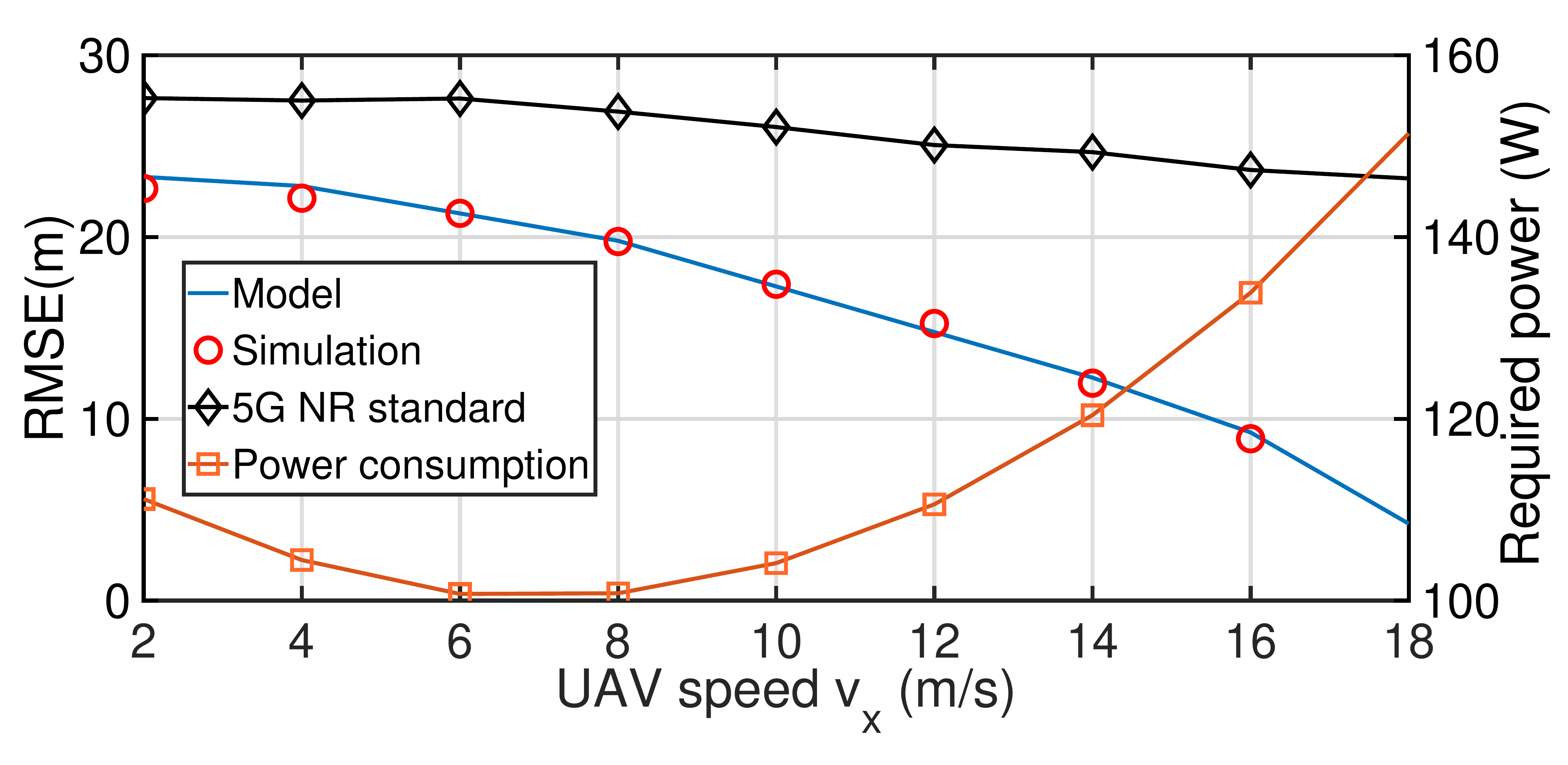}
    \caption{RMSE performance and instantaneous power consumption for different UAV speed values.}
    \label{fig:trade off}
\end{figure}

\textbf{UAV optimal speed.} The above-mentioned behavior is further confirmed by Fig.~\ref{fig:trade off}, which supports the relation between UAV tilt and speed. Again, OTFS delivers higher RMSE performance gains w.r.t. 5G-NR OFDM. Additionally, Fig.~\ref{fig:trade off} shows that the optimal UAV speed can be determined as a trade-off between ranging accuracy and power consumption. 

\change{In~\cite{Zeng2019}, for a rotary-wing with speed \(v_{x}\), the propulsion power consumption is modeled, given a constant-power transmit signal as 
\begin{align}
W\!=\!W_0\bigg(1+\frac{3v_{x}^2}{U_{tip}^2}+ \frac{d_{0}\rho sAv_{x}^3}{2W_0} \bigg)\!+\!W_{i}\bigg(\sqrt{1+\frac{v_{x}^4}{4v_{0}^4}}-\frac{v_{x}^2}{2v_{0}^2}\bigg)^{\frac{1}{2}}\hspace{-0.2cm},
\label{eq:power consumption vs v}
\end{align}
where $W_0$ and $W_i$ are constants indicating the blade profile power and the induced power in hovering status, respectively. \(U_{tip}\) is the tip speed of the rotor blade, \(v_{0}\) denotes the mean rotor induced velocity in hover, \(d_{0}\) is known as the fuselage drag ratio, \(s\) stands for the rotor solid, \(\rho\) and A are the air density and rotor disc area, respectively.} By comparing the RMSE and Power Consumption curves in Fig.~\ref{fig:trade off}, the proper speed value minimizing both of them can be selected. The proposed scheme allows not only to localize the user with higher accuracy but also to minimize the consumed power with respect to the 5G-NR standard. This aspect appears to be crucial in case of emergency situations with battery-operated equipment. 


\vspace{1mm}
\section{Conclusions}
\label{sec:conclusions}

In this work, a UAV-mounted BS is used to localize victims equipped with cellular devices in emergency situations. Specifically, we leverage on the novel OTFS modulation scheme to propose a PRACH procedure with the aim of spreading the preamble sequence in the delay-Doppler domain thereby reducing the impact of multipath fading and Doppler spread due to the UAV motion. 
We have demonstrated high ranging accuracy and, as a consequence, high target localization accuracy by means of insightful simulation results, obtained through a commercial ray-tracing simulator, namely Wireless InSite, and a mathematical tool, namely MATLAB. When 5G numerology is exploited, adapting the subcarrier space in the OTFS modulation leads to a reduced sampling distance and, in turn, to better accuracy. Lastly, we have shown the existence of a trade-off on the UAV speed to reduce the consumed power while increasing the overall localization process accuracy.

\vspace{1mm}
\section{Acknowledgements}
This work was supported by EU H2020 RISE-6G project under Grant number 101017011.
\vspace{1mm}

\bibliographystyle{ieeetr}
\bibliography{Bibliography}

\end{document}